\documentclass[12pt]{article}

\input epsf
\def\beq{\begin{eqnarray}}
\def\eeq{\end{eqnarray}}
\def\m{M_*}
\def\mpl{M_{\rm Pl}}

\def\L*{{\cal L}_*}
\def\lsim{\mathrel{\rlap{\lower3pt\hbox{\hskip0pt$\sim$}}
     \raise1pt\hbox{$<$}}}         %less than or approx. symbol
\def\gsim{\mathrel{\rlap{\lower4pt\hbox{\hskip1pt$\sim$}}
     \raise1pt\hbox{$>$}}}         %greater than or approx. symbol

\usepackage{amsmath}
\usepackage{amsfonts}

\begin{document}

\begin{titlepage}

\begin{flushright}
{NYU-TH-07/03/16}
\end{flushright}
\vskip 0.9cm

\centerline{\Large \bf Carg\`ese Lectures on Brane Induced Gravity}

\vskip 0.7cm
\centerline{\large Gregory Gabadadze}
\vskip 0.3cm
\centerline{\em Center for Cosmology and Particle Physics}
\centerline{\em Department of Physics, New York University, New York, 
NY, 10003, USA}

\vskip 1.9cm

\begin{abstract}

A brief introduction is given to the subject of brane induced gravity.
The 5D example is discussed in detail.  The 4D laws of gravity are 
obtained on a brane  embedded in an infinite 
volume extra space, where the problem of stabilization of  the 
volume modulus is absent. The theory has two classically disjoint 
branches of solutions -- the conventional and self-accelerated one. 
The conventional  branch gives  a perturbatively stable model of a 
metastable graviton, with potentially testable predictions 
within the Solar system.  The self-accelerated branch, on the other hand, 
provides an existence proof for an idea that the accelerated expansion 
of  the Universe could be due to modified gravity. The issue of 
perturbative stability of the self-accelerated branch is obscured by 
a breakdown  of the conventional perturbative expansion. 
However, a certain exact non-perturbative solution found in 
hep-th/0612016 exhibits a net negative gravitational mass,
while this mass is positive on the conventional branch.
This suggest that the self-accelerated solution must be non-perturbatively 
unstable. A proposal to overcome this  problem  in an  extension of 
the original model, that also allows for the quantum gravity scale to be 
unrestricted, is briefly discussed.

\end{abstract}

\vspace{3cm}

\end{titlepage}

\newpage

\section{Introduction}

The discovery  of Refs. \cite {Acc}  that that the present-day 
expansion of the  Universe is accelerating, has been confirmed 
by a number of subsequent efforts.
One way to parametrize the accelerated expansion is to postulate
the existence of a ``dark energy'' component in the Einstein equation:
\beq
G_{\mu\nu} = 8\pi G_N (T^{\rm matter}_{\mu\nu}+ 
T^{\rm dark~ energy}_{\mu\nu})\,,
\label{Einstein}
\eeq
were $G_{\mu\nu}=R_{\mu\nu} -g_{\mu\nu}R/2$, is the 4D Einstein tensor
of the metric $g_{\mu\nu}(x)$, and 
$T^{\rm matter}_{\mu\nu} $ and $T^{\rm dark~ energy}_{\mu\nu}$
denote the stress-tensors for matter (including dark matter) 
and dark energy, respectively; the latter has to have a negative 
enough pressure to account for the observations.

On the other hand, one can consider another logical 
possibility that the accelerated expansion 
is due to modified General Relativity (GR). 
Schematically, the modified Einstein equation 
could be written as:
\beq
G_{\mu\nu} - {\cal K}_{\mu\nu}(g,m_c) = 
8\pi G_NT^{\rm matter}_{\mu\nu}\,,
\label{MGR}
\eeq
where  ${\cal K}_{\mu\nu}(g,m_c)$ denotes a tensor that could depend 
on a metric $g$, its derivatives, as well as on other fields not present in 
GR. Moreover, ${\cal K}$  contains a dimensionful constant
$m_c\sim H_0\sim 10^{-42}~GeV$, which sets a distance/time  
scale $r_c\equiv m^{-1}_c$ at which the Newtonian potential obtained  
from (\ref {MGR}) significantly  deviates from the conventional one. 
One concrete example of ${\cal K}_{\mu\nu}(g,m_c)$ is given by 
the DGP model \cite {DGP}, and  will be discussed in detail in the 
next section.

At a first sight, there does not seem to be a difference between 
(\ref {MGR}) and (\ref {Einstein}), as the new term on the l.h.s.
of (\ref {MGR}) could be transfered to its r.h.s.  and regarded as 
the ``dark energy'' component, similar to the one  
present in (\ref {Einstein}).

In reality, however, the difference between (\ref {MGR}) and (\ref {Einstein})
is significant.  It is typically implied in (\ref {Einstein}) 
that the dark energy is either  due to cosmological constant 
or a light scalar  field. This field
forms an independent sector of the theory, the dynamics of which is 
not restricted by severe constraints that general covariance 
imposes on tensor fields.  In contrast with this, the new term in (\ref {MGR})
contains metric itself in a nontrivial way. As such it is 
highly restricted by general covariance. This gives rise to  both 
the theoretical and observational differences between models 
of ``dark energy'' and modified gravity.

It is worth emphasizing that a primary theoretical motivation 
for the  models of modified gravity is to evade 
the S.Weinberg's no-go theorem  on the ``old'' cosmological 
constant problem (CCP) \cite {Weinberg}, see  Ref. \cite {GGrev} 
for a summary of these discussions.  This attractive possibility 
still exists, in principle, in models of brane induced gravity with 
the number of space-time dimensions $D\ge 6$ \cite {DG,DGS,GS}, however, 
many  aspects  of those models are not  well-understood, 
and we won't be discussing them here. Instead we concentrate on the 
5D brane induced gravity \cite {DGP}. The latter does 
not offer a solution to the old CCP, however, it can be used as 
an example for  understanding of a new dynamics introduced 
by modified gravity.

\section{Brane induced gravity}

An explicit example of the modified Einstein equation  (\ref {MGR}) is 
provided by the DGP model \cite {DGP}. All the known  4D interactions, 
except gravity,  are  thought to be confined to a brane that is 
embedded in a 5D  infinite-volume (uncompactified) 
empty space where only gravity propagates. 
In this setup, the additional term on the l.h.s. of (\ref {MGR}) is 
provided by the 4D extrinsic curvature terms of the brane
\beq
G_{\mu\nu} - {m_c} (K_{\mu\nu} - g_{\mu\nu} K)
= 8\pi G_NT_{\mu\nu}(x)\,.
\label{junction}
\eeq
Here, $G_{\mu\nu}=R_{\mu\nu} -g_{\mu\nu}R/2$, is the 4D Einstein 
tensor for the metric that depends on both 4D coordinates $x_\mu$, 
and  the fifth coordinate $y$,   $g_{\mu\nu}(x,y)$;  
$K=g^{\mu \nu}K_{\mu \nu}$, is the trace of the 
extrinsic curvature tensor 
\beq
K_{\mu \nu}\,=\,{1\over 2N}\,\left (\partial_y g_{\mu \nu} -\nabla_\mu 
N_\nu
-\nabla_\nu N_\mu \right )\,,
\label{K}
\eeq
and $\nabla_\mu $ is a 4D covariant derivative w.r.t. the metric 
$g_{\mu \nu}(x,y)$. We introduced the {\it lapse} scalar field  $N$, and 
the {\it shift} vector field $N_\mu $ \cite {ADM}:
\beq
g_{\mu 5}\,\equiv \, N_\mu =g_{\mu\nu}N^\nu\,,~~~g_{55}\,\equiv \,N^2\,+\,
g_{\mu\nu}\,N^\mu \,N^\nu\,. 
\label{adm}
\eeq 
In the DGP model, equation (\ref {junction}) 
is accompanied by the  $\{\mu\nu \}$ equation in the bulk, and 
the $\{\mu 5\}$ and  $\{5 5\}$ equations which respectively read
\beq
G^{(5)}_{\mu\nu} =0\, ~~~{\rm for}~~~y\neq 0\,, 
\label{bulkmunu} \\
\nabla^\mu K_{\mu\nu} = \nabla_\nu K\,,
\label{mu5new} \\
R= K^2 -K_{\mu\nu}K^{\mu\nu}\,.
\label{55} 
\eeq
Here $G^{(5)}_{\mu\nu}$  denotes the 5D Einstein tensor
for the 5D metric $g_{AB}(x,y)$ (A,B=0,1,2,3,{\it 5}), and 
$g_{\mu\nu}(x,y)$  is its 4D part.
Note that the $\{\mu 5\}$ and  $\{5 5\}$ equations,
(\ref {mu5new}) and (\ref {55}),  should be satisfied in 
the bulk, $y\neq 0$ , as well as  on the brane, $y=0$.

The above set of equations can be derived by the variational principle
using the  action \cite {DGP}
\beq
S\,=\,{\mpl^2 \over 2} \,\int\,d^4x\, \sqrt{g_4}\,R(g_4)\,+
{\m^{3} \over 2}\,\int \,d^4x\,dy\,\sqrt{g_5}
\,{\cal R}(g_5)\,+{\rm surface~terms},
\label{1}
\eeq
where $g_{4\mu\nu}\equiv g_{\mu\nu}(x,0)$, and 
 $g_5$ refers to the full 5D metric; $R$ and ${\cal R}$ are 
the four-dimensional
and five-dimensional Ricci scalars respectively, and
$\m$, stands for the fundamental gravitational scale
of the bulk theory.  The brane  is located  at $y=0$ and ${\bf Z}_2$
symmetry across the brane is imposed. The 
boundary Gibbons-Hawking term should be taken into account  
to warrant the correct Einstein equations in the bulk.
The matter fields, that are also omitted here for simplicity, 
are assumed to be localized on the brane.

Another form of the action 
(\ref {1}) can be given in the ADM formalism
\beq
S\,=\,{\mpl^2 \over 2} \,\int\,d^4x\, 
\sqrt{g_4}\,R(g_4)\,+ {\m^{3} \over 2} \int \,d^4x\,dy\,\sqrt{g}
N(R+ K^2 -K_{\mu\nu}^2)\,.
\label{12}
\eeq
The above action gives rise to the equations  of motion
(\ref {junction},\ref {bulkmunu}-\ref {55}), where 
\beq
m_c \equiv {2 M^3_*\over \mpl^2}\,.
\label{mc}
\eeq
In order for $m_c$ to be of order $H_0$, we need that 
the bulk quantum gravity scale to be $\m \sim 100~MeV$. Such a low value 
of the quantum gravity scale is in no conflict with  observations, 
however, it presents an obstacle for a string theory realization of the above 
model. In the next section, we will show how a mild extension of the 
model can relax the constraint on  $\m$, in which case it could take 
an arbitrary value  below the 4D Planck mass \cite {GGnew}.

\section{Role of non-linearities}

To better understand the linearized theory
of (\ref {1}), it is instructive to consider an analogous
scalar model \cite {DGP}
\beq
-{\mpl^2 \over 2} \int d^4x  
(\partial_\mu \phi(x,0))^2 - {\m^3 \over 2}
\int d^4x dy (\partial_A \phi(x,y))^2\,,
\label{scalar1}
\eeq
where the dimensionless scalar field $\phi$ is 
to mimic the 5D graviton of the previous section. 
We impose the ${\bf Z_2}$ symmetry across
the $y=0$ boundary,  and add to the above action 
the coupling of $\phi$ to  a source $J$, also localized on the brane,    
$\int d^4x dy \delta (y) J\phi$. To obtain the junction condition, 
we integrate the equation of motion obtained  from  
(\ref {scalar1}) w.r.t. $y$, from $0^{-}$ to $0^+$.
The resulting equation written at $y=0^+$ reads:
\beq
- \partial^2_\mu \phi|_{y=0}  - m_c \partial_y\phi|_{y=0^+}= J/\mpl^2\,.  
\label{junctionscalar}
\eeq
The bulk equation is nothing but a 5D free scalar equation
\beq
\partial^A\partial_A \phi =0\,~~~{\rm for}~~y \neq 0\,.
\label{bulks}
\eeq
It is straightforward to find a solution to the bulk and junction equations.
For this we turn to the momentum space w.r.t. the 4D coordinates
while keeping the fifth dimension intact. 
The (decaying in the bulk) solution reads: 
\beq
{\tilde \phi}(p, y)\,=\, \left ( {\tilde J\over \mpl^2} \right )\, 
{{\rm exp} (-p |y|) \over p^2\,+\,m_c\,p}\,,
\label{solphi}
\eeq
where the sign ``tilde'' denotes the Fourier transformed quantities, and 
we introduced the Euclidean 4-momentum square as follows:
$ p^2\,\equiv\,p^{\mu}p_\mu \,=\,-p_0^2\,+\,p_1^2\,+ p_2^2\,+p_3^2\,\equiv
\,p_4^2\,+\,p_1^2\,+ p_2^2\,+p_3^2\,,
$ and set the notation $p\equiv \sqrt{p^2}$.

The solution (\ref {solphi}) exhibits a number of interesting
properties: 

(1) At short distances (i.e., the large momenta $p\gg m_c$)
it reduces to a 4D solution with the inverse square law, $1/r^2$, 
for the force mediated by this scalar.  
At large distances (i.e., small  momenta 
$p\ll m_c$) it  turns into a 5D solution with the $1/r^3$ force 
law.

(2) The expression (\ref {solphi}) appears to have two poles,
one at $p^2=0$ and another one at $p=\sqrt {p^2} =-m_c$. However,
the residue of the former pole is zero, consistent with the 
expectation that there is no normalizable 4D massless state in the 
spectrum. 

The second pole appears on a non-physical Riemann sheet of the complex 
$p^2$-plane (see details in Ref. \cite {GGrev}), 
and describes a resonance-like state. 
The branch-cut on the complex $p^2$-plane is 
due to the $\sqrt {p^2}$ term in (\ref {solphi}), and reflects the
presence of a continuum of the Kaluza-Klein (KK) states.

(3) Two localized sources on a brane exchange one 5D scalar state
the propagator of which can be read off (\ref {solphi}).
From the 4D perspective what is being exchanged is an infinite number
of KK states, the 4D couplings of which  are suppressed as compared to 
the ordinary 5D theory by the following factor \cite {DGKN}:
\beq
{1\over 1+(m^2/m_c^2)},
\label{couplings}
\eeq
where $m$ denotes the mass of a given KK state.
Therefore, the heavier the KK state the more its suppression. 
That is why  at distances $\lsim r_c=m_c^{-1}$ one recovers
4D Newtonian potential.

A similar calculation can be performed for the 
linearized gravitational theory
\cite {DGP}. The object of the primary interests here is 
the one-graviton exchange amplitude between two sources. The 
expression for the resulting amplitude is gauge independent and reads 
as follows: 
\beq
{\cal A}_{\rm 1-graviton}(p,\,y)\,=
\,{T^2_{1/3} \over p^2\,+\,m_c\,p}{\rm exp} (-p |y|)\,,
\label{A13}
\eeq
where 
\beq
T^2_{1/3} \,\equiv\, 8\,\pi\,G_N \left ( 
T^2_{\mu\nu}\,-\,{1\over 3}\,T\cdot T \right )\,.
\label{T1third}
\eeq
As in the scalar example, 
the pole at $p^2=0$ has zero residue, and the 
second pole in  (\ref {A13})  is on a non-physical Riemann sheet.
Therefore, the amplitude  ${\cal A}_{\rm 1-graviton}$   
describes propagation of a metastable state
with the lifetime $\sim m^{-1}_c$, which decays into a 
continuum of the KK modes.

However, there is a crucial difference from the scalar case. This has 
to do with the numerator of (\ref {A13}). In the limit 
$m_c\to 0$ the numerator does not reduce to the analogous expression
in GR. The latter takes the form:
\beq
8\,\pi\,G_N \left ( 
T^2_{\mu\nu}\,-\,{1\over 2}\,T\cdot T \right )\,.
\label{T1half}
\eeq
The difference between  (\ref {T1third}) and 
(\ref {T1half}) is due to the fact that a 5D graviton (or a 
massive graviton for that matter) propagates 5 on-shell degrees of 
freedom (helicity-2, helicity-1, and helicity-0), while the GR 
graviton propagates only 2 on-shell degrees of freedom 
(helicity-2 state). The helicity-1 state of the 5D graviton  
does  not contribute to the one-graviton exchange amplitude  
(\ref {T1third}) because of the conservation of the stress-tensor.
However, the helicity-0 state does contribute and gives rise 
to the finite difference between (\ref {T1third}) and (\ref {T1half}).
Observationally  (\ref {T1third}) is excluded! This is the essence of the 
van Dam-Veltman-Zakharov discontinuity (vDVZ) \cite {vDVZ}.

Fortunately, the vDVZ argument fails for observationally interesting 
sources. This is due to nonlinear 
interactions \cite {Arkady,DDGV}, and can easily be understood as 
follows \cite {DDGV}: the longitudinal  part of the graviton 
propagator in DGP contains terms proportional to 
\beq
{p_\mu \,p_\nu\over m_c p} \,.
\label{singDGP}
\eeq
This term does not contribute to the amplitude ${\cal A}_{\rm 1-graviton}$ 
because of conservation of the stress-tensor,
however, it does contribute already in the first  nonlinear 
correction (since the stress-tensor is only covariantly conserved
in the non-linear theory). Due to the singularity in 
$m_c$ in  (\ref {singDGP}), the perturbation 
theory  breaks down  precociously.
However, this breakdown is  an artifact of an ill-defined 
perturbative expansion -- the  known  exact solutions of the 
model have no trace of breaking \cite {DDGV}.
The perturbative expansion in powers of $G_N$ 
gets ``contaminated'' by another dimensionful 
parameter $1/m_c$, and this leads to its breakdown.

Under the circumstances, one could either
adopt a different type of expansion, e.g., an expansion 
in the small parameter $m_c$ \cite {DDGV,Gruzinov}, 
or look at exact solutions
\footnote{It is also possible to modify the theory at the linearized 
level so that the conventional perturbative expansion is well-behaved 
\cite {GGweak,Siopsis},\cite {Massimo},\cite{Smolyakov}.}.
Both of these programs have been carried out to a certain  
extent. Here we summarize the main results.

The model has one adjustable parameter -- the distance/time  
scale $r_c=m_c^{-1}$. Distributions of matter and  radiation which are 
homogeneous and  isotropic at scales $ \gsim r_c$ exhibit 
the  following properties: for distance/time 
scales $\ll r_c$ the solutions approximate General Relativity (GR)
to a  high degree of accuracy, while for scales $\gsim r_c$
they dramatically differ \cite {Cedric,DDG,DDGV}. 
Since  $r^{-1}_c\sim H_0\sim 10^{-42} {\rm GeV}$,
the deviations from GR could lead to observational  
consequences in late-time cosmology, see, e.g., 
\cite {DDG}, \cite{Landau}--\cite{Hu}.

On the other hand, sources of matter and radiation 
with typical inhomogeneity scales  less than $r_c$
have somewhat different properties. 
These are easier to discuss  for  a  Schwarzschild source --
a spherically-symmetric distribution of matter of the mass $M$ 
and radius $r_0$, such that $r_M < r_0  \ll r_c$ 
($r_M\equiv 2G_NM$  is the  Schwarzschild  radius).  
For such a source a new scale, that is a combinations of 
$r_c$ and $r_M$, emerges 
(the so-called  Vainshtein scale\footnote{A similar, but not 
exactly the same scale was discovered by Vainshtein in massive 
gravity \cite {Arkady}, hence the name.}) \cite {DDGV}: 
\beq
r_*\equiv (r_M r_c^2)^{1/3}\,.
\label{r*}
\eeq
Below this scale the predictions of the theory are in a 
good agreement with the GR results.
Above this scale, however, gravity of a compact object deviates 
substantially from the GR result. Note that $r_*$ is huge for 
typical astrophysical objects. An isolated star of 
a solar mass would have $r_* \sim 100~ pc$. However,  
if we draw a sphere of a $100~pc$ radius   
with the Sun  in its center  there will be many other starts
enclosed by that sphere.  The matter enclosed
by this sphere would have even larger  $r_*$. 
We could  draw a bigger sphere, but it will enclose 
more matter which would  yield yet larger  $r_*$ and so on.  
An isolated object which  could  potentially be separated from  
a neighboring  one  by a distance larger than its own $r_*$
is a cluster of galaxies. For typical clusters,  $r_*\sim (few~Mpc)$
is just somewhat larger than their size.

The above arguments suggest that interactions of 
isolated clusters may be different in the DGP model.  On the other hand,  
at scales beneath a few Mpc or so, there will be agreement with the GR 
results with potentially interesting small deviations.  

For simplicity, we  discuss below 
these issues for a single isolated  
Schwarzschild source. There exist 
in the literature two different (but both partial) 
solutions for the Schwarzschild 
problem in the DGP model.  The first one is based on approximate 
expansions in the $r \ll r_*$ and $r \gg r_*$ regions \cite 
{DGP,DDGV,Gruzinov} (see also \cite {Tanaka}). 
We call this set of results the perturbative  Schwarzschild (PS) solution. 
The second one \cite {GI} is a solution on the brane 
that interpolates smoothly 
from $r\ll r_*$  to $r\gg r_c\gg r_*$,  and is non-analytic in the 
either parameters  used to obtain the PS solution.  
We call this the non-perturbative 
Schwarzschild (NPS) solution. What is certain, 
is that at observable distances both solutions are in 
good agreement with the GR results, but predict 
a tiny and potentially measurable deviations from GR
\cite{DGZ,Lue1}  (see also \cite {Iorio1,GIlunar}).

It is important to understand which of 
these two solutions, if any, is physically viable. 
Since neither of the two  were  solved completely in the entire
5D space-time,  a first step to discriminate between them
would be to look closely at the 
predictions that could by tested 
observationally. This was discussed in detail in 
Ref. \cite {GIlunar}. We briefly summarize some of 
the results.

Let us  start with  the  Newton  
potential $\varphi(r)$. The result 
for $r\ll r_*$ leads:
\beq
-2\varphi = {r_M\over r}-\alpha m_c^2r^2 
\left({r_*\over r}\right)^{{3\over 2} - \beta }+\dots~,
\label{potential}
\eeq
where $\alpha = \pm \sqrt{2}$ and $\beta =0$ for the 
perturbative solution (PS), while $\alpha = \pm 0.84$ 
and $\beta = 3/2 - 2(\sqrt{3}-1)\simeq 0.04$
for the non-perturbative solution (NPS) \cite {GI}
(two signs for $\alpha$ correspond to the tow different branches
of solutions, see discussions below).

The deviation from 4D gravity at $r\ll r_*$ 
gives rise to  the additional perihelion 
precession of circular orbits \cite{DGZ,Lue1} 
(see also \cite {Iorio1} for comprehensive studies
of these and related issues). In a simplest approximation this effect 
is quantified by a fraction of the deviation of the potential $\varphi$
from its Newtonian form 
\beq
\epsilon \equiv {\Delta \varphi \over \varphi}\,.
\label{epsilon}
\eeq
This can be  used  to evaluate
an additional perihelion precession of orbits
in the Solar system \cite{DGZ,Lue1}\footnote{Note that in the 
leading order of the relativistic expansion the answer 
is given by the correction to the Newtonian potential, while 
the correction  to the $rr$ component of the metric is 
not important.}. The $\epsilon$ 
ratio is somewhat different for the non-perturbative solution (NPS)
as compared to the perturbative solution (PS) used in Refs. 
\cite{DGZ,Lue1}. This difference has been calculated 
\cite {GIlunar}:
\beq
{\epsilon_{NPS}\over \epsilon_{PS}} \simeq 0.59 
\left ({r\over r_*}\right )^{0.04}\,.
\label{eratio} 
\eeq
The perihelion precession per orbit is
\beq
\Delta \varphi = 2\pi  +  {3\pi r_M\over r} \mp {3\pi |\alpha| \over 4}
\left ( {r\over r_*}\right )^{3/2} \,\left ({r\over r_*}\right )^{0.04}\,.
\label{anprec}
\eeq
The second term on the RHS is the 
Einstein precession,
and the last term arises due to modification of gravity.
For the PS this was first calculated in Refs. 
\cite{DGZ,Lue1} ; the solution (\ref {anprec})
is written for the NPS and is somewhat different. 

For the Earth-Moon system $r\simeq 3.84 \times 10^{10}~cm$ and 
$r^{Earth}_* \simeq 6.59 \times 10^{12}~cm$; 
as a result the ratio in (\ref {eratio})
is approximately $0.48$. Therefore, the predictions of the 
non-perturbative solution for the additional perihelion precession of the
Moon is a factor of two smaller than the predictions of the 
perturbative  solution. The result of (\ref {anprec}) for the 
additional precession (the last term on the RHS) is 
$\mp 0.7 \times 10^{-12}$ (the plus sign 
for the self-accelerated branch). This is below  the current 
accuracy of $2.4 \times 10^{-11}$ \cite {lunardata}, but 
could potentially be probed in the near future \cite {adel}.

A similar calculations can be performed for the anomalous Martian 
precession \cite{DGZ,Lue1} . For the Sun-Mars system 
we use $r_{Sun-Mars}= 2.28 \times 10^{13}~ cm $   and 
$r^{Sun}_*=4.9 \times  10^{20} ~cm$.  The additional precession of 
the Mars orbit is $ \sim \mp 1.3\times 10^{-11}$, which should 
be contrasted with a potential accuracy of the Pathfinder mission  
$ \sim 9\times 10^{-11}$.

\section{The quantum gravity scale}

We have discussed in the previous section that the value 
of the 5D Planck mass $\m$ is restricted by the requirement that 
$m_c=2\m^3/\mpl^2$ be of order $H_0$, leading to the value 
of the bulk ``quantum gravity'' scale $\m\sim 100$ MeV. 

Here we discuss a slight modification of the DGP model which retains
the most of the important properties of the original theory and yet allows
to relax the constrain on the bulk quantum gravity scale $\m$.

For simplicity we first discuss this for the  scaler example.
The main idea is that the strength of the 5D kinetic term 
could depend on the $y$ coordinate, so that its value is small on the brane 
but is large off the brane. We could parametrize this as
\beq
\int dx dy F(x,y) (\partial_A \phi(x,y))^2\,,
\label{form}
\eeq
where the new scalar $F$ (a ``dilaton'') is assumed to have an $x$ independent 
profile in the $y$ direction such that $F(y\to 0)=m_c\mpl^2/4$,
while $F(y\neq 0)$ sets  the bulk quantum gravity scale, 
which is unrestricted and could be as large as $\mpl$. 

Let us perform these calculations more carefully. To account for the above
properties, we introduce an additional term into 
the action (\ref {scalar1}) which is just a opposite sign 
5D kinetic term peaked  on the brane. To make things tractable, we 
smear the brane, that is, instead of the Dirac function $\delta (y)$,
we use its regularized version  $\delta (y)\to {\bar \delta}(y) 
\equiv \pi^{-1} \varepsilon/(y^2 +\varepsilon^2)$, with $\varepsilon \to 0$.
The term that we'll be adding to (\ref {scalar1}) then reads:
\beq
{{M}^2\over 2}  \int dx dy {\bar \delta}(y) (\partial_A \phi(x,y))^2\,.
\label{scalaradd}
\eeq
With this term included the variation of the action $\delta S=0$ 
with the appropriate boundary conditions gives:
\beq
- (\mpl^2 -M^2) {\bar \delta}(y)  \partial^2_\mu \phi  -\m^3 \partial^2_\mu
\phi - \partial_y\left ((\m^3 -M^2 {\bar \delta}(y)) 
\partial_y \phi \right ) = 
J{\bar \delta}(y) \,.
\label{fjunctionscalar0}
\eeq
Next we take  the integral of both sides of this equation w.r.t. $y$ from
$-\varepsilon $ to $+\varepsilon $, and then turn to the limit 
\beq
M\to 0,~~~\varepsilon \to 0,~~~ M^2/\varepsilon \sim 
M^2 {\bar \delta} (0)\equiv {\bar M}^3> {\m^3}\,, 
\label{limit}
\eeq
where we keep ${\bar M}$ fixed, and its value  
somewhat lower than $\m$. The resulting equation reads:
\beq
-  \partial^2_\mu \phi
- {2 ( \m^3 - {\bar M}^3) \over \mpl^2}  \partial_y \phi = J/\mpl^2\,.
\label{fjunctionscalar1}
\eeq
Finally, introducing
\beq
m_c \equiv  {2 ( {  \m^3 - {\bar M}^3}) \over \mpl^2}\,,
\label{newmc}
\eeq
where the positive numerical value of $m_c$ will be tuned 
to the Hubble scale today  $m_c\sim H_0 \sim 10^{-42}$ GeV,
we get the desired junction conditions
\beq
- \partial^2_\mu \phi  - m_c \partial_y\phi = J/\mpl^2\,.  
\label{fjunctionscalar}
\eeq
Two comments.  First, 
the wrong-sign kinetic term (\ref {scalaradd}) that is peaked only 
on the brane is dominated  by 
the large positive 4D kinetic term in (\ref {scalar1}), 
proportional to $\mpl^2 > M^2$. Second, the number of adjusted parameters 
here is the same as in DGP:
In (\ref {scalar1}) one should tune the value of $\m$ such that the 
ratio $2\m^3/\mpl^2$ is of order $H_0$, while 
in the action (\ref {scalaradd}) the value of 
$\m<\mpl$ can be arbitrary, as long as 
one tunes the value of  ${\bar M}$ so that 
(\ref {newmc}) is of order $H_0$.

\vspace{0.1in}

In the case of gravity, to which we  turn now,
similar considerations can applied.
As before, we smooth out the brane by  replacing 
$\delta (y) \to {\bar \delta}(y)$,   and  think of the 5D 
EH term to have a profile due to the ``dilaton'' field
\beq
\int d^4x\,dy\,\sqrt{g_5} F(x,y){\cal R}(g_5),
\label{profile}
\eeq
such that the 5D gravitational coupling on the brane is 
strong, while it becomes weak in the bulk.

The above construction could  be parametrized by adding  
the following 
boundary (worldvolume) term to the DGP action:
\beq
-{M^2\over 2} \int \,d^4x\,dy\,{\bar \delta} (y) \sqrt{g_5}{\cal R}(g_5)=
-{M^2\over 2} \int \,d^4x\,dy\,{\bar \delta} (y) \sqrt{g}\, N 
\left (R+K^2-K_{\mu\nu}^2\right )\,,
\label{add}
\eeq
where the r.h.s. of (\ref {add}) is obtained by using the standard 
ADM decomposition. The total action in the ADM formalism reads:
\beq
S_{\rm mod}\,=\,{\mpl^2 \over 2} \,\int\,d^4x\, 
\sqrt{g_4}\,R(g_4)\,+ {\m^{3} \over 2} \int \,d^4x\,dy\,\sqrt{g}
N(R+ K^2 -K_{\mu\nu}^2)\, \nonumber \\
- {M^{2}\over 2}  \int \,d^4x\,dy\, {\bar \delta} (y)   \sqrt{g} 
N \left (R+ K^2 -K_{\mu\nu}^2 \right )\,,
\label{fdgp}
\eeq
where it is implied that the 4D EH term is also 
smeared over the same scale as the 5D term
\footnote{For the regularization of 4D and 5D EH terms, see,
\cite {DHGS} and  \cite {Massimo}, respectively.}. 
The equations of motion are 
straightforward to derive from (\ref {fdgp}).
The $\{\mu\nu\}$ equation in the bulk, and $\{\mu 5\}$ and $\{5 5\}$ equations
read  as follows:
\beq
G^{(5)}_{\mu\nu} =0\, ~~~{\rm for}~~~|y|> \epsilon \,, 
\label{fbulkmunu} \\
(\m^3 - M^2 {\bar \delta}(y))
\left (  \nabla^\mu K_{\mu\nu}  - \nabla_\nu K \right ) =0\,,
\label{fmu5new} \\
(\m^3- M^2 {\bar \delta}(y))\left (R - K^2 +K_{\mu\nu}K^{\mu\nu}\right ) =0\,.
\label{f55} 
\eeq
As in the scalar case, we  will be looking at this theory 
in the  limit (\ref {limit}).
The above equations reduce to (\ref {bulkmunu}-\ref {55}).

The Israel junction condition across the brane gets modified
because of the new term in (\ref {fdgp}).  In the limit 
(\ref {limit}) this condition reads:
\beq
G_{\mu\nu} - {2({ \m^3-  {\bar M^3})}\over \mpl^2} 
(K_{\mu\nu} - g_{\mu\nu} K) = 8\pi G_NT_{\mu\nu}(x)\,.
\label{junctionnew}
\eeq
If ${\bar M} =0$, as in (\ref {1}),  we get back the result
(\ref {junction}) with $m_c = 2\m^3/\mpl^2$. However,
${\bar M} $ does not have to be zero. For an 
arbitrary value of $\m$ we tune the value of ${\bar M}$
so that  the crossover scale (\ref {newmc}), 
which  appears in (\ref {junctionnew}),  
is adjusted to the value of the present-day Hubble scale 
$m_c\sim H_0 \sim 10^{-42}$ GeV. Hence, (\ref {junctionnew})
recovers (\ref {fjunction}).

What we have shown is that for $M\ll \mpl$
the junction condition is not modified 
as compared to DGP. The 5D gravitational constant in the 
bulk equations of (\ref {fdgp}) would change, though, if we were to consider 
sources extending into the bulk. However,
our  primary  interest is  in the sources localized on the brane, 
and for those, the new model (\ref {fdgp}) with $M\ll \mpl$.
recovers the results of DGP.

\section{Cosmology}

Let us turn to the cosmological solutions. 
To this end we consider  distributions of matter and radiation
that are homogeneous at scales $\gg r_c$. Therefore, the complications
due to non-linear dynamics outlined in the previous section do not 
apply to these sources\footnote{Although, those complications 
will be relevant to perturbations about the background 
cosmological solutions.}.

The metric is parametrized as follows:
\beq
ds^2 = - P^2(t,y) dt^2 + Q^2(t,y) \gamma_{ij} dx^i dx^j + 
\Sigma^2 (t,y) dy^2\,.
\label{metric}
\eeq
There are two branches of solutions that are labeled  
by an integer $\epsilon =\pm 1$  \cite {Cedric}:
\beq
P(t,y) = 1+\epsilon |y| { {\ddot a}\over \sqrt {{\dot a}^2 +k} }, ~~
Q(t,y) = a(t) + \epsilon |y| {\sqrt {{\dot a}^2+k} },~~~
\Sigma (t,y) =1\,.
\label{sas}
\eeq
Here we included a nonzero spatial curvature $k$. 
With this Ansatz, the Friedmann equation on the brane 
follows from (\ref {junction}), and can be expressed  
in terms of the 4D Hubble parameter $H\equiv {\dot a}/a$.
For a simplest case of a brane without matter/radiation on it, 
and with $k=0$ the equation  reads \cite {Cedric}
\beq
H^2 - \epsilon m_c|H| = 0\,.
\label{Friedmann}
\eeq
The case with $\epsilon =+1$  admits  a dS  solution 
with $H=m_c$. This is called the self-accelerated solution,
as it gives rise to the accelerated expansion of the universe 
due to modified gravity. We emphasize that 
the minus sign between the two terms in (\ref {Friedmann}) 
is guaranteed by the choice of the positive sign in front of the 
terms in  (\ref {sas}) that are proportional to $|y|$. 
If we were to  choose the latter signs to be negative, 
we would have obtained the Friedmann equation
$H^2+m_c |H|=0$, which does not admit the dS  solution.
The latter corresponds to the choice $ \epsilon =-1$, and is referred to 
as the conventional branch. There is no acceleration produced by 
modified gravity on this branch. It has to be obtained, e.g., by 
introducing a small cosmological constant on the brane. 
An interesting observation concerning this branch is the 
following \cite{Luew}: the cosmological  
expansion from a 4D point of view looks as if it had the equation of 
state parameter less than $-1$. This is because gravity turns into the  
5D regime at scales $H\sim m_c$, which implies less ``deceleration'' 
of matter due to attractive gravity, and  this appears  to a 4D 
observed as ``faster'' acceleration \cite {Luew}.

It is straightforward to introduce matter/radiation on the brane.
Following \cite {Cedric} we obtain  the Friedmann equation
\beq
H^2+{k\over a^2} = \left (\sqrt{{8\pi G_N\over 3}\rho +{m_c^2\over 4}} + 
\epsilon {m_c \over 2}\right )^2\,, 
\label{matter}
\eeq
which should be amended by  the conventional conservation equation 
for the fluid of density  $\rho$  and pressure $p$: ${\dot \rho} + 
3H(\rho +p)=0$. The latter being a result  of the 
matter stress-tensor conservation $\nabla^\mu T_{\mu\nu}=0$, which 
can be verified, e.g., by taking a covariant derivative 
of both sides of (\ref {junction}) and using  (\ref {mu5new}).

When $H\gg m_c$,  i.e., in   the early universe, eq. (\ref {matter})
reduces to the conventional Friedmann equation. However, at late times,
when  $H\sim  m_c$, the cosmological evolution changes. On the 
conventional branch ($ \epsilon =-1$) it turns into the expansion 
driven by the 5D force law, and on the self-accelerated branch 
($ \epsilon = 1$) it turns into the dS-like expansion.

\vspace{0.1in}

How about perturbations on the cosmological solutions?
We start with the conventional branch. Here, the small 
perturbations about the  Minkowski background are stable \cite {DGP}. 
However, the conventional perturbative expansion  breaks down 
near realistic sources, as was discussed in the previous section.  
As a result, the model exhibits the strongly-coupled behavior already 
at the classical level \cite {DDGV} in the domain
where the extrinsic curvature square is $\gsim m_c^2$.
The same applies to the self-accelerated branch.
Because of this, the question of stability of the 
self-accelerated solution \cite {Luty} -- \cite{KaloperRuth}, 
which happens to be in the strongly coupled regime due to its 
curvature, becomes difficult 
to address within the perturbative approach 
\cite {DGI,Dvali}\footnote{One could look at this breakdown in terms 
of the scalar  ``conformal'' mode. In massive gravity this mode 
decouples from the rest of the modes 
in a certain limit \cite {AGS}. In the DGP model, however, 
such a decoupling does not take place \cite {GIDec}.}.

Luckily, certain non-linear solutions have been  found.  
This is a case for a Schwarzschild-like solution, 
for which the 4D metric was  exactly obtained \cite {GI}, 
and for the Domain Wall solution for  which the full 5D 
metric was found in Ref. \cite {DGPR}.  In both cases, 
the mass (tension) of the solution gets screened by gravitational 
effects, and  these sources on the {self-accelerated background} 
look as if they had a {\it negative} net 5D mass 
(tension)\footnote{In contrast with this, screening of 
the similar sources on the conventional branch of DGP,  leaves them with  
positive 5D mass (tension) \cite {GI,DGPR}.}. 
This suggests that the self-accelerated  background  
may not be problem-free in the full non-linear theory. 

What is a root-cause of this behavior?  
The self-accelerated solution exists  only for a certain choice of 
the sign of the extrinsic curvature, 
and this choice is such that it requires a growing metric
in the bulk. For instance, in a simplest spatially-flat 
case, the full 5D metric of the self-accelerated solution takes 
the form \cite {Cedric}:
\beq
ds^2 = \left (1 +  H|y|\right )^2 \left \{ - dt^2 + e^{2Ht}
d{\vec x}^2 \right \}   
+  dy^2\,,
\label{metricSA}
\eeq
where $y$ is the 5th coordinate  and $H$ denotes the dS expansion 
rate of the 4D worldvolume (the latter is labeled here 
by Cartesian coordinates $(t, {\vec x})$, and we use the 5D coordinate 
system in which the brane is located at $y=0$.). 
The unusual feature of the above  metric is that 
it grows in the bulk,  even though the 
worldvolume metric is that of dS space. A linearly growing metric,
similar to (\ref {metricSA}), would have been produced by a {\it negative}
tension 3-brane\footnote{This could be in, e.g.,  
5D Minkowski or Anti de Sitter (AdS) space-time.}. 
The  growing metric (\ref {metricSA}) imprints 
its ``negative'' effects on the brane worldvolume through the 
extrinsic curvature, giving rise to the 
solutions mentioned in the previous paragraph.

Can this problem be cured? It turns out that one can  modify  
the DGP equations in such a way that the new system  admits a background 
that is equivalent to the self-accelerated solution on the 4D brane, 
however, differs from it in the bulk.  

The new solution that we will discuss takes the form:
\beq
ds^2 = \left (1 -  H|y|\right )^2 \left \{ - dt^2 + e^{2Ht} d{\vec x}^2 
\right \} +  dy^2\,.
\label{metricFSA}
\eeq
In order to obtain this solution one  needs
to flip the sign in front of  the extrinsic curvature term 
in one of the DGP equations, keeping the rest of the equations 
intact.  Below  will discuss
how such equations can be obtained by modifying  
the DGP action. 

The solution (\ref {metricFSA}) is formally identical 
to that for a  3-brane endowed with a positive 4D cosmological constant 
(brane tension) which is  embedded in 5D empty space in 5D GR \cite {KLinde}.
However, unlike the latter, the worldvolume expansion 
in the present case (\ref {metricFSA})  is due to modified gravity, 
while the 4D cosmological constant is set to zero. 
This difference is what is responsible 
for the modified Friedmann equation, and distinct 
cosmological evolution on the self-accelerated background. 

The bulk space in  both (\ref {metricSA}) and (\ref {metricFSA})
is locally equivalent to 5D Minkowski space. In the chosen coordinate system 
the solution (\ref {metricFSA}) encounters the Rindler horizon at 
$|y|=H^{-1}$. However, an  analytic 
continuation beyond this point  can be performed 
by employing new coordinates. In that 
coordinate system the brane (with closed spatial sections) 
can be regarded as a  4D dS bubble that is  first contracting 
and  then re-expanding in 5D Minkowski space.

\vspace{0.1in}

The proposal of Ref. \cite {GGnew} is to add a new term 
on the brane worldvolume such that  the 
sign in front of the second term on 
the l.h.s. of (\ref {junction}) would flip. In other words, we introduce 
a new set of equations in which  (\ref {junction}) 
is replaced by:
\beq
G_{\mu\nu}\, + \, {m_c} (K_{\mu\nu} - g_{\mu\nu} K)+\gamma \Sigma_{\mu\nu}
= 8\pi G_NT_{\mu\nu}(x)\,,
\label{fjunction}
\eeq
while all the other equations (\ref {bulkmunu}),(\ref {mu5new}) and 
(\ref {55}) remain intact. Here, in a simplest case  
$\Sigma_{\mu\nu}\equiv (G_{\mu\nu}- {1\over 2}g_{\mu\nu}
(K^2 -K^2_{\alpha\beta})- 2(K^\alpha_\mu K_{\nu\alpha} -K K_{\mu\nu}))$,
and the small coefficient  $\gamma\lsim 1$ is in general 
nonzero for a regularized brane width. An important property of the tensor 
$\Sigma_{\mu\nu}$ is that it equals to $G_{\mu\nu}$  on the self-accelerated 
solution given below, while it contributes in a non-trivial way 
to perturbations about it.

The action functional that gives rise to this new set of equations 
happens to be the one that we have already discussed
in the previous section (\ref {fdgp}). The only difference is that in
the junction condition (\ref {junctionnew}) we need to flip the sign of 
the coefficient in front of the extrinsic curvature terms. 
This can be achieved if we choose ${\bar M} > \m$ and 
set
\beq
m_c \equiv  {2({\bar M}^3 -\m^3)\over \mpl^2}\,.
\label{minusmc}
\eeq

The metric for the self-accelerated solution of the new 
system of equations (\ref {fjunction}, \ref {bulkmunu}--\ref {55})
reads:
\beq
P(t,y) = 1 - |y| {{\ddot a}\over \sqrt {{\dot a}^2+k}   },
~~~ Q(t,y) = a(t) - |y| \sqrt {{\dot a}^2+k },
~~~\Sigma (t,y) =1\,,
\label{fsas}
\eeq
where we have chosen a negative sign in front of the terms proportional 
to $|y|$. Let us now see how the change of the positive signs 
in the metric (\ref {sas}) into the negative signs in 
(\ref {fsas}) changes the value of the 
extrinsic curvature evaluated at $y=0^+$.
On the solution (\ref {fsas}), $N_\mu=0$, $N=1$, and 
$K_{\mu\nu} = \partial_y g_{\mu\nu}/2$.  Hence, at $y=0^+$ 
the components of the extrinsic curvature tensor evaluated on 
the solution (\ref {fsas}) equal to {\it minus}  the 
corresponding components evaluated on (\ref {sas}).  
Therefore, substitution of  (\ref {sas})
into  (\ref {junction})  would give the same 
equation as the substitution of (\ref {fsas}) into (\ref {fjunction}).
The corresponding Friedmann equation on the {\it empty} brane,
which now follows from  (\ref {fjunction}) instead of (\ref {junction}), 
reads:
\beq
H^2 + {k\over a^2}= m_c \sqrt {H^2 + {k\over a^2}}\,. 
\label{Fr}
\eeq
For $k=0$ this coincides with (\ref {Friedmann})
and gives the spatially-flat  dS solution with $H=m_c$ (\ref {metricFSA}). 
For general $k$  the solutions are: 
\beq
ds^2 = \left (1- H|y|\right )^2 \left \{ - dt^2 + a^2(t)
(d\chi^2 + S^2_k(\chi) d \Omega^2)  \right \}   +  dy^2\,,
\label{metricSol}
\eeq
where $H=m_c$ and  $k=-1,0,1$ corresponds to the open, flat and closed spatial 
slicing of 4D dS space, for which  $S_k(\chi)= {\rm sinh} \chi,  
\chi, {\rm sin} \chi$, respectively\footnote{There are 
two other solutions to (\ref {Fr}). For $k=0$ one finds the  $H=0$
flat solution. For $k=-1$ one finds the Milne solution $a(t)=t$.}.

The solution  (\ref {metricSol})  should satisfy all the bulk 
equations (\ref {bulkmunu} -- \ref {55}), since in the bulk it is 
locally equivalent to Minkowski space. We  checked by direct 
substitution that (\ref {metricSol}) solves Eqs. 
(\ref {bulkmunu} -- \ref {55}) too. The matter/radiation density
can also be introduced as  described above. The Freedman equation 
coincides with (\ref {matter}) with $\epsilon$ set to 1.
 
It still remains to be shown  that the modified model described by 
(\ref {fdgp})  gets read of all the negative mass states 
that may appear in the self-accelerated solution, this  
question will be discussed in \cite {GGnew1}.   
I just point out that an additional benefit of 
the new term in (\ref {fdgp}) is that it allows to relax the constraint on the 
bulk gravity scale. The latter can take an arbitrary value
below $\mpl$. This opens a window for a possible string theory 
realization of this model, or its  $D>5$ counterparts \cite {DG,GS}
(for earlier proposals see \cite {Ignat,Pierre}).

\subsection*{Acknowledgments}

I would like to thank the organizers of the Carg\`ese 2007 School,
especially  Laurent Baulieu, Pierre Vanhove, Paul 
Windy, and Elena Gianolio. The work is supported by NASA grant 
NNGG05GH34G and NSF grant 0403005.

\end{document}